\documentclass{article}
\usepackage{spconf}

\usepackage{algorithm,algorithmic,amsmath,amssymb,amsthm,bbm,cite,color,enumitem,flushend,graphicx,microtype,setspace,subcaption,url}
\usepackage{hyperref}
\usepackage[USenglish]{babel}
\usepackage[utf8]{inputenc}
\usepackage[T1]{fontenc}

\usepackage{pgfplots,tikz}
\usetikzlibrary{arrows}
\pgfplotsset{compat=1.17}

\newcommand{\h}{\mathbf{h}}

\newcommand{\p}{\mathbf{p}}

\renewcommand{\r}{\mathbf{r}}

\newcommand{\x}{\mathbf{x}}
\newcommand{\y}{\mathbf{y}}
\newcommand{\z}{\mathbf{z}}

\newcommand{\0}{\mathbf{0}}

\newcommand{\A}{\mathbf{A}}

\newcommand{\C}{\mathbf{C}}

\renewcommand{\H}{\mathbf{H}}
\newcommand{\I}{\mathbf{I}}

\renewcommand{\P}{\mathbf{P}}

\newcommand{\V}{\mathbf{V}}

\newcommand{\X}{\mathbf{X}}
\newcommand{\Y}{\mathbf{Y}}
\newcommand{\Z}{\mathbf{Z}}

\newcommand{\setC}{\mathcal{C}}

\newcommand{\setE}{\mathcal{E}}

\newcommand{\setN}{\mathcal{N}}

\newcommand{\setQ}{\mathcal{Q}}

\newcommand{\setS}{\mathcal{S}}

\newcommand{\Compl}{\mbox{$\mathbb{C}$}}

\newcommand{\argmin}{\operatornamewithlimits{argmin}}
\newcommand{\blkdiag}{\mathrm{blkdiag}}

\newcommand{\Diag}{\mathrm{Diag}}

\newcommand{\Exp}{\mathbb{E}}
\newcommand{\herm}{\mathrm{H}}
\renewcommand{\Im}{\mathrm{Im}}

\renewcommand{\Re}{\mathrm{Re}}
\newcommand{\sgn}{\mathrm{sgn}}
\newcommand{\tr}{\mathrm{tr}}
\newcommand{\tran}{\mathrm{T}}

\newtheorem{theorem}{Theorem}

\usepackage{fancyhdr}
\fancypagestyle{firstpage}{
  
  \fancyfoot[C]{{\footnotesize \begin{singlespace} \textcopyright 2023 IEEE. Personal use of this material is permitted. Permission from IEEE must be obtained for all other uses, in any current or future media, including reprinting/republishing this material for advertising or promotional purposes, creating new collective works, for resale or redistribution to servers or lists, or reuse of any copyrighted component of this work in other works.
\end{singlespace}}}}

\title{Multi-User Data Detection in Massive MIMO with 1-Bit ADCs}
\name{Amin Radbord, Italo Atzeni, and Antti Tölli\thanks{This work was supported by the Academy of Finland (318927 6G Flagship, 336449 Profi6, and 348396 HIGH-6G).}}
\address{Centre for Wireless Communications, University of Oulu, Finland \\
Emails: \{amin.radbord, italo.atzeni, antti.tolli\}@oulu.fi}

\begin{document}
\ninept

\maketitle

\thispagestyle{firstpage}

\begin{abstract}
We provide new analytical results on the uplink data detection in massive multiple-input multiple-output systems with 1-bit analog-to-digital converters. The statistical properties of the soft-estimated symbols (i.e., after linear combining and prior to the data detection process) have been previously characterized only for a single user equipment (UE) and uncorrelated Rayleigh fading. In this paper, we consider a multi-UE setting with correlated Rayleigh fading, where the soft-estimated symbols are obtained by means of maximum ratio combining based on imperfectly estimated channels. We derive a closed-form expression of the expected value of the soft-estimated symbols, which allows to understand the impact of the specific data symbols transmitted by the interfering UEs. Building on this result, we design efficient data detection strategies based on the minimum distance criterion, which are compared in terms of symbol error rate and complexity.
\end{abstract}


\section{Introduction} \label{sec:intro}

Increasing the capacity of beyond-5G wireless systems will require exploiting the wide bandwidths available in the THz spectrum (0.3--3~THz) \cite{Raj20}. This calls for massive multiple-input multiple-output (MIMO) arrays at the transmitter and/or at the receiver to compensate for the strong pathloss and penetration loss therein. In this regard, fully digital architectures provide highly flexible wideband beamforming and large-scale spatial multiplexing \cite{Lu14}. However, this approach requires a sacrifice in the resolution of the analog-to-digital/digital-to-analog converters (ADCs/DACs), since their power consumption scales linearly with the sampling rate and exponentially with the number of quantization bits \cite{Li17,Jac17a,Atz21b}. Remarkably, fully digital architectures with low-resolution ADCs (even down to 1 bit) can significantly outperform their hybrid analog-digital counterparts in terms of spectral and energy efficiency \cite{Rot18}. In this respect, 1-bit ADCs/DACs are particularly attractive as they are the simplest and least power consuming data conversion devices \cite{Li17,Mol17}. Such a coarse quantization is suitable with very large bandwidths, for which high-order modulations may not be needed.

There is a vast literature on 1-bit quantized massive MIMO, ranging from performance analysis (e.g., \cite{Li17,Mol17}) to data detection (e.g., \cite{Cho16,Atz22}) and precoding (e.g., \cite{Sax17,Jac17b,Jac19}). In this paper, we broaden prior analytical studies on the uplink data detection in massive MIMO systems with 1-bit ADCs. The statistical properties of the soft-estimated symbols (i.e., after linear combining and prior to the data detection process) have been characterized in our prior works \cite{Atz22,Atz21a,Abd23} only under a simplified system model with a single user equipment (UE) and uncorrelated Rayleigh fading. However, uncorrelated channel models cannot account for the sparse scattering at high frequencies. In this paper, we consider a more general and realistic multi-UE setting with correlated Rayleigh fading. We assume that the base station (BS) adopts maximum ratio combining (MRC) based on imperfectly estimated channels, where the channel estimation is carried out via the Bussgang linear minimum mean squared error (BLMMSE) estimator \cite{Li17}. In this context, we derive a closed-form expression of the expected value of the soft-estimated symbols, which is relevant to understand the impact of the specific data symbols transmitted by the interfering UEs. This result is exploited to design efficient data detection strategies based on the minimum distance criterion, which are compared in terms of symbol error rate (SER) and complexity.

\section{System Model} \label{sec:SM}

Let us consider a single-cell massive MIMO system where a BS with $M$ antennas serves $K$ single-antenna UEs in the uplink. We use $\H \triangleq [\h_{1}, \ldots, \h_{K}] \in \Compl^{M \times K}$ to denote the uplink channel matrix. Considering a general correlated Rayleigh fading channel model, we have $\h_{k} \sim \setC \setN (\0, \C_{\h_{k}}),~\forall k$, where $\C_{\h_{k}} \in \Compl^{M \times M}$ is the channel covariance matrix of UE~$k$. Furthermore, we define $\h \triangleq \mathrm{vec} (\H) \in \Compl^{M K}$ and, accordingly, we have $\h \sim \setC \setN (\0, \C_{\h})$, with $\C_{\h} \triangleq \blkdiag (\C_{\h_{1}}, \ldots ,\C_{\h_{K}}) \in \Compl^{M K \times M K}$. For simplicity, and without loss of generality, we assume that all the UEs are subject to the same SNR $\rho$ during both the channel estimation and the uplink data transmission (see also \cite{Atz22}). Each BS antenna is connected to two 1-bit ADCs, one for the in-phase and one for the quadrature component of the received signal. In this context, we introduce the 1-bit quantization function $Q(\cdot) : \Compl^{A \times B} \rightarrow \setQ$, with $\setQ \triangleq \sqrt{\frac{\rho K + 1}{2}} \{ \pm 1 \pm j \}^{A \times B}$ and \cite{Jac17a,Atz22}
\begin{align} \label{eq:Q}
Q(\X) \triangleq \sqrt{\frac{\rho K + 1}{2}} \big( \sgn( \Re[\X]) + j \, \sgn (\Im[\X]) \big).
\end{align}

\subsection{Data Detection} \label{sec:SM_1}

Let $\x \triangleq [x_{1}, \ldots, x_{K}]^{\tran} \in \Compl^{K}$ denote the data symbol vector comprising the data symbols transmitted by the UEs. We assume that $\x \in \setS^{K}$, where $\setS \triangleq \{ s_{1}, \ldots, s_{L} \}$ represents the set of the $L$ possible data symbols. For instance, $\setS$ may correspond to the 16-QAM constellation, as considered in Section~\ref{sec:NR}. During the uplink data transmission, all the UEs simultaneously transmit their data symbols, and the signal received at the BS is given by
\begin{align} \label{eq:y}
\y \triangleq \sqrt{\rho} \H \x + \z \in \Compl^{M}
\end{align}
where $\z \in \Compl^{M}$ is the additive white Gaussian noise (AWGN) vector with i.i.d. $\setC \setN (0,1)$ elements. The BS observes the quantized signal
\begin{align} \label{eq:r}
\r \triangleq Q(\y) \in \Compl^{M}
\end{align}
where the scaling factor in \eqref{eq:Q} is such that the variance of $\r$ coincides with that of $\y$. Then, the BS obtains a soft estimate of $\x$ via linear combining as
\begin{align}
\label{eq:x_hat_1} \hat{\x} & \triangleq [\hat{x}_{1}, \ldots, \hat{x}_{K}]^{\tran} \\
\label{eq:x_hat_2} & = \V^{\herm} \r \in \Compl^{K}
\end{align}
where $\V \in \Compl^{M \times K}$ is the combining matrix. Finally, the data detection process maps each soft-estimated symbol in \eqref{eq:x_hat_1} to one of the possible data symbols in $\setS$.

\subsection{Channel Estimation} \label{sec:SM_2}

The combining matrix $\V$ used in \eqref{eq:x_hat_2} is designed based on the estimated channels. In this paper, the channel estimation is carried out via the BLMMSE estimator \cite{Li17}, which is the state-of-the-art linear estimator with low-resolution ADCs and reduces to the well-known minimum mean squared error estimator in the absence of quantization. Let $\P \triangleq [\p_{1}, \ldots, \p_{K}] \in \Compl^{\tau \times K}$ denote the pilot matrix, where $P_{u,k}$ represents its ($u,k$)th element and $\tau$ is the pilot length. Assuming $\tau \geq K$, orthogonal pilots among the UEs, and $|P_{u,k}|^{2} = 1,~\forall u,k$, we have $\P^{\herm} \P = \tau \I_{K}$. During the channel estimation, all the UEs simultaneously transmit their pilots, and the signal received at the BS is given by
\begin{align} \label{eq:Yp}
\Y_{\mathrm{p}} \triangleq \sqrt{\rho} \H \P^{\herm} + \Z_{\mathrm{p}} \in \Compl^{M \times \tau}
\end{align}
where $\Z_{\mathrm{p}} \in \Compl^{M \times \tau}$ is the AWGN matrix with i.i.d. $\setC \setN (0,1)$ elements. At this stage, we vectorize \eqref{eq:Yp} as
\begin{align}
\y_{\mathrm{p}} & \triangleq \mathrm{vec} (\Y_\mathrm{p}) \\
& = \sqrt{\rho} \bar{\P}^{*} \h + \z_\mathrm{p} \in \Compl^{M \tau}
\end{align}
with $\bar{\P} \triangleq \P \otimes \I_{M} \in \Compl^{M \tau \times M K}$ and $\z_\mathrm{p} \triangleq \mathrm{vec}(\Z_\mathrm{p}) \in \Compl^{M \tau}$. Furthermore, we define
\begin{align}
\C_{\y_\mathrm{p}} & \triangleq \Exp [\y_{\mathrm{p}} \y_{\mathrm{p}}^{\herm}] \\
& = \rho\ \bar{\P}^{*} \C_\h \bar{\P}^\tran + \I_{M \tau} \in \Compl^{M \tau \times M \tau}
\end{align}
and
\begin{align}
\A_\mathrm{p} \triangleq \sqrt{\frac{2}{\pi}(\rho K + 1)} \Diag(\C_{\y_\mathrm{p}})^{-\frac{1}{2}} \in \Compl^{M \tau \times M \tau}.
\end{align}
The BS observes the quantized signal 
\begin{align} \label{rp}
\r_\mathrm{p} & \triangleq Q(\y_\mathrm{p}) \in \Compl^{M \tau}
\end{align}
and obtains the estimate of $\h$ via the BLMMSE estimator as
\begin{align}
\hat{\h} & \triangleq \sqrt{\rho} \C_\h \bar{\P}^\tran \A_\mathrm{p} \C_{\r_\mathrm{p}}^{-1} \r_\mathrm{p} \in \Compl^{M K}
\end{align}
with $\C_{\r_\mathrm{p}} \triangleq \Exp [\r_\mathrm{p} \r_\mathrm{p}^\herm]$. Finally, the estimate of $\H$ is given by $\hat{\H} \triangleq [\hat{\h}_{1}, \ldots, \hat{\h}_{K}]$, with
\begin{align} \label{eq:h_hat_k}
\hat{\h}_k \triangleq \sqrt{\rho} \C_{\h_k} \bar{\p}_{k}^\tran \A_\mathrm{p} \C_{\r_\mathrm{p}}^{-1} \r_\mathrm{p} \in \Compl^{M}
\end{align}
and $\bar{\p}_{k} \triangleq \p_{k} \otimes \I_{M} \in \Compl^{M \tau \times M}$.

\section{Data Detection Analysis} \label{sec:main}

In our prior work \cite{Atz22}, we focused on a single-UE setting with uncorrelated Rayleigh fading and characterized the statistical properties of the soft-estimated symbols. In this paper, we consider a more general and realistic multi-UE setting with correlated Rayleigh fading and provide a closed-form expression of the expected value of the soft-estimated symbols. This result, presented in Section~\ref{sec:main_1}, allows to understand the impact of the specific data symbols transmitted by the interfering UEs. Furthermore, it can be exploited to design efficient data detection strategies based on the minimum distance criterion, as described in Section~\ref{sec:main_2}.

\subsection{Expectation of the Soft-Estimated Symbols} \label{sec:main_1}

As in \cite{Atz22}, we consider that the MRC receiver is adopted at the BS. Hence, the combining matrix is given by $\V = \hat{\H}$ and the soft-estimated symbol for UE~$k$ can be expressed as $\hat{x}_{k} = \hat{\h}_{k}^{\herm} \r$ (cf. \eqref{eq:x_hat_2}), with $\hat{\h}_{k}$ and $\r$ given in \eqref{eq:h_hat_k} and \eqref{eq:r}, respectively. Let us define $\C_{\r \r_\mathrm{p}} \triangleq \Exp [\r \r_\mathrm{p}^{\herm}]$, which represents the cross-correlation matrix between the quantized signals received during the uplink data transmission and the channel estimation. Moreover, we introduce the function $\Omega(x) \triangleq \frac{2}{\pi} \arcsin(x)$ and the following preliminary definitions:
\begin{align}
\alpha_{m} & \triangleq \bigg[ \rho \sum_{k=1}^{K} \C_{\h_k} + \I_M \bigg]_{m,m}, \\
\beta_{m} & \triangleq \bigg[ \rho \sum_{k=1}^{K} \C_{\h_k} |x_{k}|^{2} + \I_M \bigg]_{m,m}, \\
\zeta_{m,n,u,v} & \triangleq \frac{\rho}{\sqrt{\alpha_{m} \alpha_{n}}} \bigg[ \sum_{k=1}^{K} \C_{\h_{k}}^{\tran} P_{u,k} P_{v,k}^{*} \bigg]_{m,n}, \\
\eta_{m,n,u} & \triangleq \frac{\rho}{\sqrt{\alpha_{n} \beta_{m}}} \bigg[ \sum_{k=1}^{K} \C_{\h_{k}} x_{k} P_{u,k} \bigg]_{m,n}.
\end{align}
The following theorem provides a closed-form expression of the expected value of the soft-estimated symbol for UE~$k$ for a given data symbol vector $\x$. This is denoted by $\mathsf{E}_{k} \triangleq \Exp [\hat{x}_{k}]$, where the expectation is taken over $\H$, $\z$, and $\z_\mathrm{p}$.

\begin{theorem} \label{thm:main}
Assuming that the MRC receiver is adopted at the BS, for a given data symbol vector $\x$, the expected value of the soft-estimated symbol for UE~$k$ is given by
\begin{align} \label{eq:Ex}
\mathsf{E}_{k} = \sqrt{\rho} \tr( \C_{\r_\mathrm{p}}^{-1} \A_\mathrm{p} \bar{\p}_{k}^{*} \C_{\h_k} \C_{\r \r_\mathrm{p}})
\end{align}
where the ($(u-1)M+m,(v-1)M+n$)th element of $\C_{\r_\mathrm{p}}$ can be written as in \eqref{eq:Crp} at the top of the next page and the ($m,(u-1)M+n$)th element of $\C_{\r \r_\mathrm{p}}$ can be written as
\begin{align}
\nonumber [\C_{\r \r_\mathrm{p}}]_{m,(u-1)M+n} & = (\rho K + 1) \Big( \Omega \big( \Re [\eta_{m,n,u}] \big) \\
\label{eq:Crrp} & \phantom{=} \ + j \, \Omega \big( \Im [\eta_{m,n,u}] \big) \Big).
\end{align}
\end{theorem}

\begin{figure*}[t!]
\begin{align}
\label{eq:Crp} [\C_{\r_\mathrm{p}}]_{(u-1)M+m,(v-1)M+n} & = \begin{cases}
\rho K + 1, & \quad \textrm{if}~m=n~\textrm{and}~u=v, \\
(\rho K + 1) \Big( \Omega \big( \Re [\zeta_{m,n,u,v}] \big) - j \, \Omega \big( \Im [\zeta_{m,n,u,v}] \big) \Big), & \quad \textrm{otherwise}
\end{cases}
\end{align}
\hrulefill
\end{figure*}

The proof of Theorem~\ref{thm:main} follows similar (and more involved) steps as in \cite[App.~D]{Atz22}. It is omitted due to the space limitations and will be provided in the extended version of this paper. The expression in \eqref{eq:Ex} for UE~$k$ clearly depends on the specific data symbols transmitted by the interfering UEs. In the following, we build on Theorem~\ref{thm:main} to design efficient data detection strategies, which will be compared in Section~\ref{sec:NR}.

\subsection{Data Detection Strategies} \label{sec:main_2}

In this section, we exploit Theorem~\ref{thm:main} and the minimum distance criterion to map each soft-estimated symbol in \eqref{eq:x_hat_1} to one of the possible data symbols in $\setS$. In this respect, we present three data detection strategies: 1) exhaustive, 2) heuristic, and 3) genie-aided data detection. In the following, we use $l_{k}^{\star} \in \{1, \ldots, L\}$ to denote the index of the detected symbol for UE~$k$.
\begin{itemize}[leftmargin=*]
\item[$\bullet$] \textit{Strategy~1: exhaustive data detection.} This strategy uses the statistical information of the interfering UEs to detect the symbol for the target UE. Let $\setE_{k} \triangleq \{ \mathsf{E}_{k}, \forall \x \in \setS^{K} \}$ denote the set of the expected values of the soft-estimated symbols for UE~$k$ obtained from all the possible data symbol vectors, with $|\setE_{k}| = L^{K}$. 
The soft-estimated symbol for UE~$k$ is mapped to one of the elements in $\setE_{k}$ as
\begin{align}
\mathsf{E}_{k}^{\star} = \argmin_{\mathsf{E}_{k} \in \setE_{k}} |\hat{x}_{k} - \mathsf{E}_{k}|
\end{align}
from which $l_{k}^{\star}$ is readily obtained. This strategy amounts to performing an exhaustive search over all the $L^{K}$ possible values of $\mathsf{E}_{k}$ in \eqref{eq:Ex}. Hence, its complexity increases exponentially with $K$.

\item[$\bullet$] \textit{Strategy~2: heuristic data detection.} This strategy considers the expected values of the soft-estimated symbols for the target UE averaged over all the possible data symbols transmitted by the interfering UEs. Let $\x_{-k} \triangleq [x_{1}, \ldots, x_{k-1}, x_{k+1}, \ldots x_{K}]^{\tran} \in \Compl^{K-1}$ and let $\setE_{k,l} \triangleq \{ \mathsf{E}_{k} : x_{k} = s_{l},~\forall \x_{-k} \in \setS^{K-1} \} \subset \setE_{k}$ be the set containing the elements in $\setE_{k}$ corresponding to $x_{k} = s_{l}$, with $|\setE_{k,l}| = L^{K-1}$. Furthermore, let us define
\begin{align} \label{eq:E_kl}
\bar{\mathsf{E}}_{k,l} \triangleq \frac{1}{L^{K-1}} \sum_{t \in \setE_{k,l}} t
\end{align}
which represents the average of the expected values of the soft-estimated symbols for UE~$k$ when $x_{k} = s_{l}$ (see the green markers in Fig.~\ref{fig:scatter_K=3_tau=3_61}). Then, the index of the detected symbol for UE~$k$ is obtained as
\begin{align}
l_{k}^{\star} = \argmin_{l \in \{1, \ldots, L\}} |\hat{x}_{k} - \bar{\mathsf{E}}_{k,l}|.
\end{align}
This strategy can be seen as a low-complexity, heuristic version of \textit{Strategy~1}, which reduces the size of the search space from $L^K$ to $L$.

\item[$\bullet$] \textit{Strategy~3: genie-aided data detection.} This strategy is obtained from \textit{Strategy~1} by assuming that a genie instantaneously provides the data symbols transmitted by the interfering UEs to detect the symbol for the target UE. Hence, for UE~$k$, $\x_{-k}$ is assumed to be perfectly known, which reduces the size of the search space from $L^K$ to $L$. Evidently, this strategy cannot be implemented in practice and is considered only to evaluate how the knowledge of the data symbols transmitted by the interfering UEs impacts the data detection performance for the target UE.
\end{itemize}

Other practical data detection strategies (e.g., resulting from combining \textit{Strategy~1} and \textit{Strategy~3} above) will be considered in the extended version of this paper.

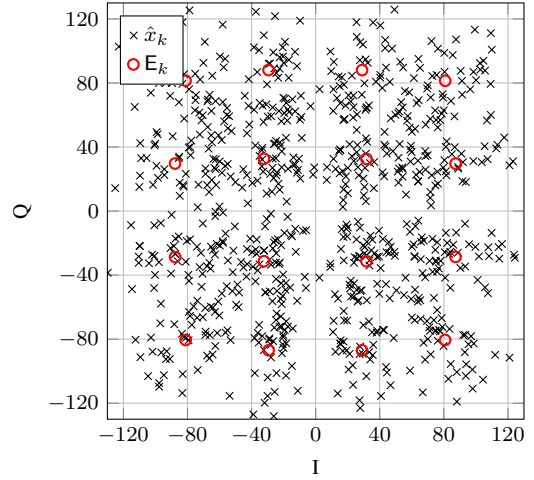
\begin{figure}[t!]
\centering
\pgfdeclarelayer{background}
\pgfdeclarelayer{foreground}
\pgfsetlayers{background,main,foreground}

\begin{tikzpicture}

\begin{axis}[
	width=0.4\textwidth,
	height=0.4\textwidth,
	xmin=-130, xmax=130,
	ymin=-130, ymax=130,
	axis equal,
    xlabel={I},
    ylabel={Q},
    xlabel near ticks,
	ylabel near ticks,
    xtick={-120,-80,-40,0,40,80,120},
    ytick={-120,-80,-40,0,40,80,120},
    legend pos=north west,
    legend cell align=left,
	grid=both,
	x label style={font=\footnotesize},
	y label style={font=\footnotesize},
	ticklabel style={font=\footnotesize},
	clip marker paths=true,
]

\addplot[only marks, black, mark=x, thin]
table[x=real_shat, y=imag_shat, col sep=comma]
{Figures/Data/fig1_s_hat_tau=61.txt};
\addlegendentry{\footnotesize{$\hat{x}_{k}$}}

\addplot[only marks, red, mark=o, thick]
table[x=real_EX, y=imag_EX, col sep=comma]
{Figures/Data/fig1_mean_and_expected_tau=61.txt};
\addlegendentry{\footnotesize{$\mathsf{E}_{k}$}}

\end{axis}

\end{tikzpicture}
\caption{3-UE scenario ($K = 3$) with $\tau = 61$: soft-estimated symbols (black markers) and their expected values (red markers) for UE~$1$ when $x_{2} = \frac{1}{\sqrt{10}}(-3 + j \, 3)$ and $x_{3} = \frac{1}{\sqrt{10}}(-3 + j)$.} \label{fig:scatter_K=3_tau=3_fixed}
\end{figure}

\section{Numerical Results} \label{sec:NR}

In this section, we utilize Theorem~\ref{thm:main} and the data detection strategies described in Section~\ref{sec:main_2} to evaluate the impact of the specific data symbols transmitted by the interfering UEs in a massive MIMO system with 1-bit ADCs. We assume that the BS is equipped with $M = 128$ antennas and adopts the MRC receiver. The set of data symbols $\setS$ corresponds to the 16-QAM constellation, i.e., $\setS = \frac{1}{\sqrt{10}} \big\{ \pm 1 \pm j, \pm 1 \pm j \, 3, \pm 3 \pm j, \pm 3 \pm j \, 3 \big\}$, which is normalized such that $\frac{1}{L} \sum_{l=1}^{L} |s_{l}|^{2} = 1$. We consider a 2-UE scenario ($K = 2$) and a 3-UE scenario ($K = 3$). The channel covariance matrices are generated according to the one-ring channel model \cite{Yin13} with angular spread of $30^{\circ}$ for each UE and angular separation between the UEs of $120^{\circ}$ and $60^{\circ}$ for the 2-UE and 3-UE scenarios, respectively. All the UEs are subject to the same (normalized) pathloss, such that $\tr(\C_{\h_{k}}) = M,~\forall k$; unless otherwise stated, we consider $\rho = 0$~dB. The channels are estimated as described in Section~\ref{sec:SM_2} with orthogonal pilots chosen as Zadoff-Chu sequences, which are widely adopted in the 4G LTE and 5G NR standards \cite{Hyd17}; unless otherwise stated, we fix $\tau = 61$.

\begin{figure}[t!]
\centering
\pgfdeclarelayer{background}
\pgfdeclarelayer{foreground}
\pgfsetlayers{background,main,foreground}

\begin{tikzpicture}

\begin{pgfonlayer}{background}
\begin{axis}[
	width=0.4\textwidth,
	height=0.4\textwidth,
	xmin=-130, xmax=130,
	ymin=-130, ymax=130,
	axis equal,
    xlabel={I},
    ylabel={Q},
    xlabel near ticks,
	ylabel near ticks,
    xtick={-120,-80,-40,0,40,80,120},
    ytick={-120,-80,-40,0,40,80,120},
    legend pos=north west,
    legend cell align=left,
	grid=both,
	x label style={font=\footnotesize},
	y label style={font=\footnotesize},
	ticklabel style={font=\footnotesize},
	clip marker paths=true,
]

\addplot[only marks, red, mark=o, thick]
table[x=Ex_real, y=Ex_imag, col sep=comma]
{Figures/Data/fig2_expected_tau=61.txt};
\addlegendentry{\footnotesize{$\mathsf{E}_{k}$}}

\coordinate (zoom_coord) at (axis cs:125,-125);
\draw[black] (axis cs:78.2,78.2) rectangle (axis cs:78.4,78.4) {};
\begin{scope}[>=latex]
\draw[->] (axis cs:78.3,78.2) -- (axis cs:58,-9) {};
\end{scope}

\end{axis}
\end{pgfonlayer}

\begin{pgfonlayer}{foreground}
\begin{axis}[
	axis background/.style={fill=white},
	at={(zoom_coord)},
	anchor={outer south east},
	width=0.2\textwidth,
	height=0.2\textwidth,
	xmin=78.2, xmax=78.4,
	ymin=78.2, ymax=78.4,
	axis equal,
    xtick={78.2,78.3,78.4},
    ytick={78.2,78.3,78.4},
	ticklabel style={font=\tiny},
	restrict x to domain=78.2:78.4,
	restrict y to domain=78.2:78.4,
	name=zoom_axis,
]

\addplot[only marks, red, mark=o, thick]
table[x=Ex_real, y=Ex_imag, col sep=comma]
{Figures/Data/fig2_expected_tau=61.txt};

\end{axis}
\end{pgfonlayer}

\begin{pgfonlayer}{main}
\draw[black,fill=white] (zoom_axis.outer south west) rectangle (zoom_axis.outer north east);
\end{pgfonlayer}

\end{tikzpicture}
\caption{2-UE scenario ($K = 2$) with $\tau=61$: expected values of the soft-estimated symbols for UE~$1$ when UE~$2$ transmits all the possible data symbols from $\setS$.} \label{fig:scatter_K=2_tau=61}
\end{figure}
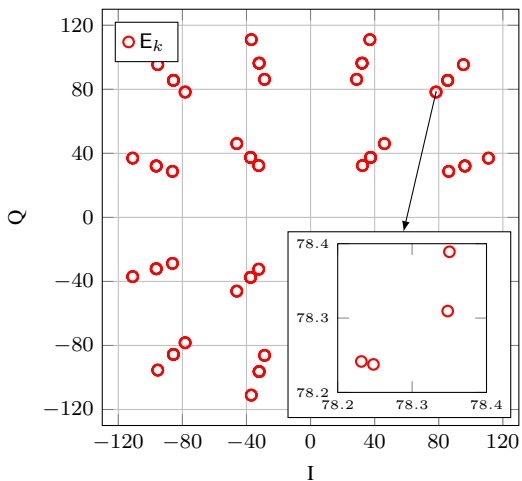

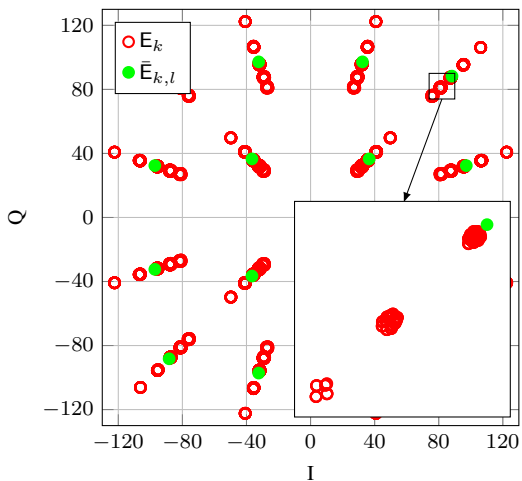
\begin{figure}[t!]
\centering
\pgfdeclarelayer{background}
\pgfdeclarelayer{foreground}
\pgfsetlayers{background,main,foreground}

\begin{tikzpicture}

\begin{pgfonlayer}{background}
\begin{axis}[
	width=0.4\textwidth,
	height=0.4\textwidth,
	xmin=-130, xmax=130,
	ymin=-130, ymax=130,
	axis equal,
    xlabel={I},
    ylabel={Q},
    xlabel near ticks,
	ylabel near ticks,
    xtick={-120,-80,-40,0,40,80,120},
    ytick={-120,-80,-40,0,40,80,120},
    legend pos=north west,
    legend cell align=left,
	grid=both,
	x label style={font=\footnotesize},
	y label style={font=\footnotesize},
	ticklabel style={font=\footnotesize},
	clip marker paths=true,
]

\addplot[only marks, red, mark=o, thick]
table[x=Var1, y=Var2, col sep=comma]
{Figures/Data/fig4_EX_tau=61.txt};
\addlegendentry{\footnotesize{$\mathsf{E}_{k}$}}

\addplot[only marks, green, mark=*, thick]
table[x=Var1, y=Var2, col sep=comma]
{Figures/Data/fig4_average_of_EX_tau=61.txt};
\addlegendentry{\footnotesize{$\bar{\mathsf{E}}_{k,l}$}}

\coordinate (zoom_coord) at (axis cs:125,-125);
\draw[black] (axis cs:74,74) rectangle (axis cs:90,90) {};
\begin{scope}[>=latex]
\draw[->] (axis cs:82,74) -- (axis cs:58,9) {};
\end{scope}

\end{axis}
\end{pgfonlayer}

\begin{pgfonlayer}{foreground}
\begin{axis}[
	axis background/.style={fill=white},
	at={(zoom_coord)},
	anchor={outer south east},
	width=0.25\textwidth,
	height=0.25\textwidth,
	xmin=74, xmax=90,
	ymin=74, ymax=90,
	axis equal,
	ticks=none,
	restrict x to domain=74:90,
	restrict y to domain=74:90,
]

\addplot[only marks, red, mark=o, thick]
table[x=Var1, y=Var2, col sep=comma]
{Figures/Data/fig4_EX_tau=61.txt};

\addplot[only marks, green, mark=*, thick]
table[x=Var1, y=Var2, col sep=comma]
{Figures/Data/fig4_average_of_EX_tau=61.txt};

\end{axis}
\end{pgfonlayer}

\end{tikzpicture}
\caption{3-UE scenario ($K = 3$) with $\tau=61$: expected values of the soft-estimated symbols (red markers) and their mean values for UE~$1$ (green markers) when UE~$2$ and UE~$3$ transmit all the possible data symbols from $\setS$.} \label{fig:scatter_K=3_tau=3_61}
\end{figure}

Considering the 3-UE scenario, Fig.~\ref{fig:scatter_K=3_tau=3_fixed} provides the scatter plot of the soft-estimated symbols for UE~$1$ when this transmits all the possible data symbols and the interfering UEs transmit fixed data symbols. Here, the soft-estimated symbols (black markers) originate from independent channel and AWGN realizations. We observe that, for each data symbol transmitted by UE~$1$, the mean value of the soft-estimated symbols is in agreement with the corresponding expected value obtained as in \eqref{eq:Ex} (red markers).

Considering the 2-UE scenario, Fig.~\ref{fig:scatter_K=2_tau=61} depicts the expected values of the soft-estimated symbols for UE~$1$ when both UEs transmit all the possible data symbols. Note that there are $16^2 = 256$ different pairs of data symbols transmitted by the two UEs, each corresponding to a different value of $\mathsf{E}_{k}$ in \eqref{eq:Ex}. However, in Fig.~\ref{fig:scatter_K=2_tau=61}, only $3 \times 16$ points can be clearly distinguished, which implies that there is significant overlap among many of the $256$ values of $\mathsf{E}_{1}$. This stems from the fact that there are three different amplitude levels in the 16-QAM constellation and, for a given data symbol transmitted by UE~$1$, the data symbols with the same amplitude transmitted by UE~$2$ produce nearly the same value of $\mathsf{E}_{1}$ (as shown in the zoomed window).

Fig.~\ref{fig:scatter_K=3_tau=3_61} extends the insights of Fig.~\ref{fig:scatter_K=2_tau=61} to the 3-UE scenario. Here, there are $16^3 = 4096$ different triplets of data symbols transmitted by the three UEs, each corresponding to a different value of $\mathsf{E}_{k}$ in \eqref{eq:Ex}. As in the 2-UE scenario, for a given data symbol transmitted by UE~$1$, the data symbols with the same amplitude transmitted by UE~$2$ and UE~$3$ produce nearly the same value of $\mathsf{E}_{1}$. Interestingly, the dispersion of such values of $\mathsf{E}_{k}$ reduces as the pilot length increases, since the channel estimates become more accurate. Moreover, the green markers in Fig.~\ref{fig:scatter_K=3_tau=3_61} correspond to the average of the expected values of the soft-estimated symbols for UE~$1$ when this transmits a specific data symbol (see \textit{Strategy~2} in Section~\ref{sec:main_2}).

\begin{figure}[t!]
\centering
\vspace{-2mm} \begin{tikzpicture}

\begin{axis}[
	width=0.45\textwidth,
	height=0.4\textwidth,
	xmin=-10, xmax=50,
	ymin=0.05, ymax=0.4,
    xlabel={$\rho$ [dB]},
    ylabel={SER},
    xlabel near ticks,
	ylabel near ticks,
    xtick={-10,0,10,20,30,40,50},
    ytick={0.05,0.1,0.15,0.2,0.25,0.3,0.35,0.4},
    yticklabels={0,0.05,0.1,0.15,0.2,0.25,0.3,0.35,0.4},
	log ticks with fixed point,
	ymode=log,
    legend pos=north east,
    legend cell align=left,
	grid=both,
	x label style={font=\footnotesize},
	y label style={font=\footnotesize},
	ticklabel style={font=\footnotesize},
]


\addplot[line width=1pt, red]
table [x=rho_dB, y=SER2, col sep=comma] 
{Figures/Data/SER_120deg_with_Av_EX.txt};
\addlegendentry{\footnotesize{Strategy 1: exhaustive}}

\addplot[line width=1pt, green]
table [x=rho_dB, y=SER4, col sep=comma] 
{Figures/Data/SER_120deg_with_Av_EX.txt};
\addlegendentry{\footnotesize{Strategy 2: heuristic}}
    
\addplot[line width=1pt, blue]
table [x=rho_dB, y=SER3, col sep=comma] 
{Figures/Data/SER_120deg_with_Av_EX.txt};
\addlegendentry{\footnotesize{Strategy 3: genie-aided}}
    
\end{axis}

\end{tikzpicture}
\caption{2-UE scenario ($K = 2$) with $\tau=31$: SER versus the SNR obtained with the data detection strategies presented in Section~\ref{sec:main_2}.} \label{fig:SER}
\vspace{-1mm}
\end{figure}
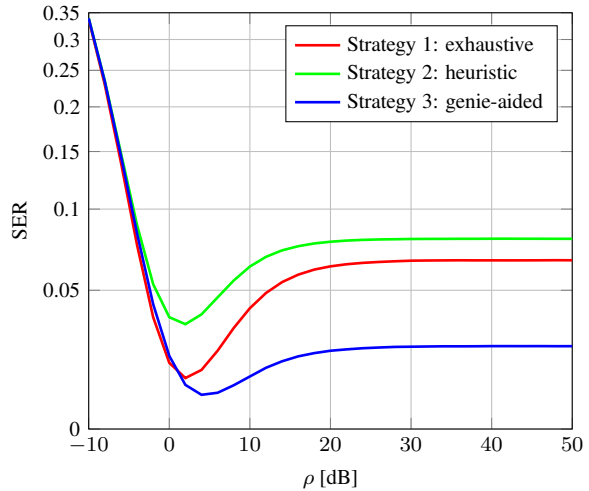

Lastly, we evaluate the performance of the data detection strategies described in Section~\ref{sec:main_2}, which are based on Theorem~\ref{thm:main} and the minimum distance criterion. Considering the 2-UE scenario, Fig.~\ref{fig:SER} plots the SER obtained with the different data detection strategies as a function of the SNR $\rho$, with $\tau = 31$. In this context, the SER is computed by averaging over $10^4$ independent channel and AWGN realizations, and considering all the possible data symbols. As in the single-UE analysis in \cite{Atz22}, we observe that the SER curves feature an optimal SNR operating point: at low SNR, the AWGN is dominant; at high SNR, the soft-estimated symbols corresponding to data symbols with the same phase are hardly distinguishable. In between these regimes, the right amount of AWGN produces a useful scrambling of the 1-bit quantized signals at the $M$ antennas. As expected, \textit{Strategy~1} outperforms \textit{Strategy~2}, since the latter corresponds to a heuristic single-UE data detection after averaging over all the possible data symbols transmitted by the interfering UE (see the green markers in Fig.~\ref{fig:scatter_K=3_tau=3_61}). Nonetheless, even \textit{Strategy~2} yields an acceptable performance at the optimal SNR operating point and beyond, partly due to the additional useful scrambling produced by the interfering UE. Furthermore, \textit{Strategy~3} outperforms all the other strategies, but \textit{Strategy~1} is remarkably close at the optimal SNR operating point.

\section{Conclusions}

We studied the uplink data detection in massive MIMO system with 1-bit ADCs considering a multi-UE setting with correlated Rayleigh fading, where the soft-estimated symbols are obtained by means of MRC based on imperfectly estimated channels. We derived a closed-form expression of the expected value of the soft-estimated symbols, which is relevant to understand the impact of the specific data symbols transmitted by the interfering UEs. Building on this result, we designed efficient data detection strategies based on the minimum distance criterion, which are compared in terms of SER and complexity. Motivated by the superior performance of the genie-aided data detection, which requires the knowledge of the data symbols transmitted by the interfering UEs, future work will focus on developing practical methods for joint data detection.


\bibliographystyle{IEEEbib}
\bibliography{refs_abbr,refs}

\begin{thebibliography}{10}

\bibitem{Raj20}
N.~{Rajatheva}, I.~{Atzeni}, E.~{Björnson}, et~al.,
\newblock ``White paper on broadband connectivity in {6G},''
  http://jultika.oulu.fi/files/isbn9789526226798.pdf, 2020.

\bibitem{Lu14}
L.~{Lu}, G.~Y. {Li}, A.~L. {Swindlehurst}, A.~{Ashikhmin}, and R.~{Zhang},
\newblock ``An overview of massive {MIMO}: {B}enefits and challenges,''
\newblock {\em IEEE J. Sel. Topics Signal Process.}, vol. 8, no. 5, pp.
  742--758, 2014.

\bibitem{Li17}
{Y. Li}, {C. Tao}, {G. Seco-Granados}, {A. Mezghani}, {A. L. Swindlehurst}, and
  {L. Liu},
\newblock ``Channel estimation and performance analysis of one-bit massive
  {MIMO} systems,''
\newblock {\em IEEE Trans. Signal Process.}, vol. 65, no. 15, pp. 4075--4089,
  2017.

\bibitem{Jac17a}
S.~{Jacobsson}, G.~{Durisi}, M.~{Coldrey}, U.~{Gustavsson}, and C.~{Studer},
\newblock ``Throughput analysis of massive {MIMO} uplink with low-resolution
  {ADCs},''
\newblock {\em IEEE Trans. Wireless Commun.}, vol. 16, no. 6, pp. 1304--1309,
  2017.

\bibitem{Atz21b}
I.~{Atzeni}, A.~{Tölli}, and G.~{Durisi},
\newblock ``Low-resolution massive {MIMO} under hardware power consumption
  constraints,''
\newblock in {\em Proc. Asilomar Conf. Signals, Syst., and Comput. (ASILOMAR)},
  2021.

\bibitem{Rot18}
{K. Roth}, {H. Pirzadeh}, {A. L. Swindlehurst}, and {J. Nossek},
\newblock ``A comparison of hybrid beamforming and digital beamforming with
  low-resolution {ADCs} for multiple users and imperfect {CSI},''
\newblock {\em IEEE J. Sel. Topics Signal Process.}, vol. 12, no. 3, pp.
  484--498, 2018.

\bibitem{Mol17}
C.~{Mollén}, J.~{Choi}, E.~G. {Larsson}, and R.~W. {Heath},
\newblock ``Uplink performance of wideband massive {MIMO} with one-bit
  {ADCs},''
\newblock {\em IEEE Trans. Wireless Commun.}, vol. 16, no. 1, pp. 87--100,
  2017.

\bibitem{Cho16}
{J. Choi}, {J. Mo}, and {R. W. Heath},
\newblock ``Near maximum-likelihood detector and channel estimator for uplink
  multiuser massive {MIMO} systems with one-bit {ADCs},''
\newblock {\em IEEE Trans. Wireless Commun.}, vol. 64, no. 5, pp.
  2005--–2018, 2016.

\bibitem{Atz22}
I.~{Atzeni} and A.~{Tölli},
\newblock ``Channel estimation and data detection analysis of massive {MIMO}
  with 1-bit {ADCs},''
\newblock {\em IEEE Trans. Wireless Commun.}, vol. 21, no. 6, pp. 3850--3867,
  2022.

\bibitem{Sax17}
{A. Saxena}, {I. Fijalkow}, and {A. L. Swindlehurst},
\newblock ``Analysis of one-bit quantized precoding for the multiuser massive
  {MIMO} downlink,''
\newblock {\em IEEE Trans. Signal Process.}, vol. 65, no. 17, pp. 4624--4634,
  2017.

\bibitem{Jac17b}
S.~{Jacobsson}, G.~{Durisi}, M.~{Coldrey}, T.~{Goldstein}, and C.~{Studer},
\newblock ``Quantized precoding for massive {MU-MIMO},''
\newblock {\em IEEE Trans. Commun.}, vol. 65, no. 11, pp. 4670--4684, 2017.

\bibitem{Jac19}
S.~{Jacobsson}, G.~{Durisi}, M.~{Coldrey}, and C.~{Studer},
\newblock ``Linear precoding with low-resolution {DACs} for massive
  {MU-MIMO-OFDM} downlink,''
\newblock {\em IEEE Trans. Wireless Commun.}, vol. 18, no. 3, pp. 1595--1609,
  2019.

\bibitem{Atz21a}
I.~{Atzeni} and A.~{Tölli},
\newblock ``Uplink data detection analysis of 1-bit quantized massive {MIMO},''
\newblock in {\em Proc. IEEE Int. Workshop Signal Process. Adv. in Wireless
  Commun. (SPAWC)}, 2021.

\bibitem{Abd23}
D.~Abdelhameed, K.~Umebayashi, I.~Atzeni, and A.~Tölli,
\newblock ``Enhanced signal detection and constellation design for massive
  {SIMO} communications with 1-bit {ADCs},''
\newblock {\em IEEE Access}, vol. 11, pp. 11749–--11765, 2023.

\bibitem{Yin13}
{H. Yin}, {D. Gesbert}, {M. Filippou}, and {Y. Liu},
\newblock ``A coordinated approach to channel estimation in large-scale
  multiple-antenna systems,''
\newblock {\em IEEE J. Sel. Areas Commun.}, vol. 31, no. 2, pp. 264--273, 2013.

\bibitem{Hyd17}
M.~{Hyder} and K.~{Mahata},
\newblock ``{Zadoff-Chu} sequence design for random access initial uplink
  synchronization in {LTE}-like systems,''
\newblock {\em IEEE Trans. Wireless Commun.}, vol. 16, no. 1, pp. 503--511,
  2017.

\end{thebibliography}

\end{document}